# Superconductivity in 1111-type CaFeAsF$_{1-x}$H$_x$ induced by selective hydrogen elimination

Taku Hanna,[1] Yoshinori Muraba,[1] Satoru Matsuishi,[3] and Hideo Hosono[1,2,3,*]

[1]Materials and Structures Laboratory, Tokyo Institute of Technology, 4259-S2-13 Nagatsuta-cho, Midori-ku, Yokohama 226-8503, Japan

[2]Frontier Research Center, Tokyo Institute of Technology, 4259-S2-13 Nagatsuta-cho, Midori-ku, Yokohama 226-8503, Japan

[3]Materials Research Center for Element Strategy, Tokyo Institute of Technology, 4259-S2-16 Nagatsuta-cho, Midori-ku, Yokohama 226-8503, Japan

*Corresponding author: Hideo Hosono

Materials and Structures Laboratory, Tokyo Institute of Technology

4259 Nagatsuta-cho, Midori-ku, Yokohama 226-8503, Japan

TEL +81-45-924-5359

FAX +81-45-924-5339

E-mail: hosono@msl.titech.ac.jp

**Abstract**

The difference in thermal stability of F$^-$ and H$^-$ in 1111-type iron based superconductors, allows selective hydrogen elimination from non-superconductive CaFeAsF$_{0.8}$H$_{0.2}$ by thermal annealing. The analyzed chemical composition of the resulting samples indicates that incorporated hydrogen was selectively eliminated by thermal annealing at 553 K for 72 hours. The resulting hydrogen-eliminated sample shows bulk superconductivity with $T_c$ = 29 K. This technique may be used for indirect electron doping for $Ae$FeAeF ($Ae$; alkali-earth metal) iron based superconductor described by CaFeAsF$_{1-x}$H$_x$ → CaFeAsF$_{1-x}$ + $xe^-$ + 1/2$x$H$_2$↑.

**Main Text:**

In the study of recently discovered iron-arsenide superconductors,[1] carrier doping has been shown to be a powerful technique to induce superconductivity. The idealized chemical composition of iron-arsenides feature iron with a formal charge of +2 and $d^6$ electron configuration, these compounds are generally not superconducting and have antiferromagnetic ordering within the Fe square lattices. Elemental substitution and defect formation supply either electrons or holes to the Fe 3$d$ bands that may suppress the magnetic ordering and induce superconductivity.[2-6] The 1111 type iron-arsenides including $Ln$FeAsO ($Ln$ = rare-earth metals) and $Ae$FeAsF ($Ae$ = alkali-earth metals) are classes of iron-pnictide superconductors. $Ln$FeAsO compounds are composed of a FeAs conducting layer and a $Ln$O blocking layer and have been systemically studied by a number of research groups as prototypical iron-based superconductors. The highest $T_c$ of iron-based superconductors (~56 K)[7] have been obtained by electron doping of $Ln$FeAsO compounds through partial substitution of oxygen with fluoride/hydride ions[8-13] and generation of oxygen vacancies.[14-17] The substitution of iron with other transition metals ($TM$) including Co, Ni and Rh, also induces superconductivity.[18-21] However, the $T_c$ induced from iron-substitution is much lower than that of oxygen site modification, indicating indirect electron doping without disrupting the FeAs layer is a better way of realizing superconductivity with a high-$T_c$. Whereas $Ln$FeAsO compounds have been studied by these carrier modification techniques, $Ae$FeAsF compounds have only been studied with respect to rare-earth substitution of $Ae$ sites, giving $T_c$ values as high as 56 K,[22,23] or $TM$-substitution of iron sites.[24-26] Superconductivity arising from substitution and defect formation induced by introduction of fluoride has not yet been observed. This is likely because of the difficulty of indirect electron doping through elemental substitution in the $Ae$F blocking layer. Therefore, alternative doping techniques are required to modify the blocking layer of $Ae$FeAsF to enable indirect electron doping.

Recently, we reported the high-pressure synthesis of 1111-type CaFeAsH, which contains hydride anions H$^-$ with a structure analogous to that of the fluoride F$^-$-containing compound CaFeAsF. Hydride and fluoride ions have similar charge and anionic radius, allowing the successful synthesis of a solid solution of CaFeAsF$_{1-x}$H$_x$.[27] However, the H-substituted samples decompose at ~873 K at ambient pressure, while CaFeAsF is stable up to 1173 K. Considering the difference in the thermal stabilities of F$^-$ and H$^-$ substituted compounds, it may be possible to selectively eliminate hydrogen to form anionic vacancies, donating electrons to the FeAs-layers.

In this paper, we describe thermal annealing of CaFeAsF$_{0.8}$H$_{0.2}$ and the electromagnetic properties of the resulting CaFeAsF$_{0.8}$H$_{0.2}$ following H-elimination. The composition of the annealed CaFeAsF$_{0.8}$H$_{0.2}$ sample was determined using an electron probe microanalyzer (EPMA) and thermogravimetry and mass spectroscopy (TG-MS) revealing that selective hydrogen elimination (CaFeAsF$_{1-x}$H$_x$ → CaFeAsF$_{1-x}$+ $xe^-$+1/2$x$H$_2$$^\uparrow$) was achieved through low temperature thermal annealing at 553 K for 72 h under a helium gas flow (99.999%, 200 mL/min). In addition, we observed bulk superconductivity with $T_c^{onset}$ = 29 K from the hydrogen-eliminated CaFeAsF$_{0.8}$H$_{0.2}$, which is higher than the $T_c^{onset}$ reported for $Ae$FeAsF (25 K) fabricated using direct doping techniques.[24-26]

A belt-type anvil cell was employed for the high pressure synthesis of $CaFeAsF_{0.8}H_{0.2}$. CaAs and $Fe_2As$ were prepared from their corresponding metals (Ca: 99.99% Aldrich, Fe: 99.9% Kojyundo, As: 99.9999% Kojyundo) and $CaH_2$ were synthesized by heating calcium metal in a $H_2$ atmosphere. All starting materials and precursors for the synthesis were prepared in a glove-box (Miwa Mfg.) filled with purified Ar gas ($H_2O$, $O_2$ < 1 ppm). The mixture of starting materials was placed into a BN capsule. Following the internal hydrogen source technique developed by Fukai and Okuma,[28, 29] $LiAlH_4$ (98%, Tokyo Kasei) was also placed in the capsule with a BN separator to provide a supplementary hydride source. The capsules were heated at 1273 K for 1 hour under a pressure of 2 GPa. To selectively eliminate hydrogen, the as-synthesized sample was ground into a powder using an agate mortar and annealed at 553 K for 72 h under a helium gas flow (99.999%, 200 mL/min) in the sample chamber of the TG-MS apparatus. Crystalline phases of the as-prepared and annealed samples were identified by powder X-ray diffraction (XRD) with Cu Kα radiation ($\lambda = 0.154056$ nm) using a Bruker diffractometer model D8 ADVANCE (Cu rotating anode). A Rietveld analysis of XRD patterns was performed using TOPAS code.[30] To investigate the hydrogen content in samples, TG-MS measurements were performed by using a Bruker AXS TG-DTA/MS9610 equipped with a gas feed port to inject standard $H_2$ gas into the sample chamber. The samples were heated up to 1073 K with a heating rate of 20 K/min under the helium gas flow. Hydrogen molecules released from a samples were ionized and detected by a quadrupole mass spectrometer as ions with $m/z = 2$. The composition of other elements (Ca:Fe:As:F) excluding hydrogen was determined using an electron-probe microanalyzer (EPMA, JEOL, Inc. model JXA-8530F) equipped with a field emission electron gun and wavelength dispersive X-ray detectors. The micrometer scale compositions within the main phase were probed in five focal points, and the results were averaged. After annealing, the powdered samples were pressed into a pellet at 300 MPa by using cold isostatic pressing, then 4-probe resistivity of pellet (ρ) and magnetic susceptibility (χ) of the powdered sample were measured in the temperature range of 2–300 K, using a physical properties measurements system (PPMS, Quantum Design) with a vibrating sample magnetometer attachment.

Figure 1 shows powder XRD patterns of the as-prepared and annealed CaFeAsF$_{0.8}$H$_{0.2}$, respectively. In the as-prepared samples, major peaks were indexed to the tetragonal ZrCuSiAs-type structure (space group: $P4/nmm$) and minor peaks to CaF$_2$ and CaO phases (< 5 wt%). Rietveld analysis indicated that the observed XRD pattern could be explained by a model structure composed of alternate stacking layers of FeAs and CaF$_{1-x}$H$_x$, with estimated lattice constants of $a$ = 0.3958(2) nm and $c$ = 0.8504(9) nm corresponding to 23% hydrogen substitution ($x$ ~ 0.23). These findings are comparable to our previous reports on CaFeAsF$_{1-x}$H$_x$ ($0 \leq x \leq 1$).[27] For annealed sample, more than 68% of the main phase remained, with contributions from impurities including CaF$_2$ (<5 wt%), CaO (<6 wt%) and FeAs (<21 wt%) phases arising from decomposition of CaFeAsF$_{0.8}$H$_{0.2}$. The lattice constants of the remaining main phase were estimated to be $a$ = 0.3896(7) nm and $c$ = 0.8669(2) nm, respectively. Although there were no clear changes of the lattice constant along the $a$-axis compared with as-prepared sample, the $c$-axis expanded by 1.6%. Notably the $c$-axis value is also 0.8% longer than that of CaFeAsF ($c$ = 0.8596 nm). Therefore lattice expansion along the $c$ axis cannot be explained by formation of CaFeAsF through decomposition of CaFeAsF$_{0.8}$H$_{0.2}$. (CaFeAsF$_{0.8}$H$_{0.2}$ → 0.8CaFeAsF + 0.2Ca(CaO) + 0.2FeAs + 0.1H$_2$↑).

The amount of hydrogen incorporated into both the as-prepared and annealed samples was evaluated by TG-MS, as shown in Figure 2. In both samples, emission of H$_2$ was observed in the range of 473–873 K. Two or more peaks were observed in effusion profiles of both samples, which are considered to be related to the surface area of grounded powder. During the TG-MS measurement up to 1073 K, the sample was decomposed into a mixture of CaFe$_2$As$_2$, FeAs and unknown phases. The amount of H$_2$ released from the as-prepared sample was measured as 3.08 mmol/g from the integrated intensity of the $m/z$ = 2 mass peak. This amount of H$_2$ was evaluated to be close to that expected from decomposition of the CaFeAsF$_{0.8}$H$_{0.2}$ phase (CaFeAsF$_{1-x}$H$_x$ →1/2(1−$x$) CaF$_2$ + 1/2CaFe$_2$As$_2$ + $x$Ca(CaO) + 1/2$x$ H$_2$↑, 2.91 mmol/g), further suggesting the formation of CaFeAsF$_{0.8}$H$_{0.2}$. Compared with the as-prepared sample, that of the annealed sample showed one tenth the mass loss of the as-prepared sample, indicating the most of hydrogen in the CaFeAsF$_{0.8}$H$_{0.2}$ was eliminated by annealing at 553 K for 72 h under a He flow (99.999%, 200 mL/min). In addition to the estimation of hydrogen content measured by TG-MS, elemental composition ratios for Ca, Fe, As, F, and O were estimated by EPMA and normalized by iron content shown in Table I. The elemental composition ratios determined by EPMA show deviation from standard samples (~5%) and from influence of impurities of oxide phases (~9%), respectively. The relationships between the hydride and fluoride compositions of the annealed and as-prepared samples are consistent with the XRD results and support the hydrogen substitution model. Despite the clear decrease of hydrogen content from 0.24 to 0.03, no other compositional differences were observed between the as-prepared and annealed samples. The results of elemental analysis using TG-MS and EPMA indicate that hydrogen was eliminated selectively from partially-hydrogen-substituted $CaFeAsF_{1-x}H_x$ samples by thermal annealing, and fluorine vacancies were formed in the 1111-type $Ae$FeAsF.

Figure 3(a) shows the $\chi$-$T$ curve of the H-eliminated $CaFeAsF_{0.8}H_{0.2}$ at 10 Oe. A sudden drop of $\chi$ is observed at ~29 K in the annealed sample owing to a superconducting transition. The shielding volume fraction (SVF) of the annealed sample excluding the volume of impurity phases was estimated at 31% from $M$-$H$ curves at 2 K (inset, FIG. 3 (a)). Although the XRD pattern of annealed sample indicates 32 % of the main phase decomposed during thermal annealing, all impurity phases identified were non-superconducting CaO, $CaF_2$ and FeAs, indicating that bulk superconductivity with a SVF of 31% was observed in hydrogen-eliminated $CaFeAsF_{0.8}H_{0.2}$.

Figure 3(b) shows the $\rho$-$T$ curves of the as-prepared sample and the H-eliminated pellet sample. The as-prepared samples exhibits shoulder-like behavior in the $\rho$-$T$ curves around ~100 K ($T_{anom}$). Similar anomalies above 100 K were also observed in both parent compounds of 1111-type and 122-type iron arsenides and correspond to a crystallographic phase transition from a tetragonal to orthorhombic lattice, although these samples did not show superconductivity. However, a sudden drop of resistivity at $T^*$ ~29 K was observed in the $\rho$-$T$ curve of the H-eliminated sample, which corresponds to a $\chi$ drop indicating a resistivity drop in H-eliminated $CaFeAsF_{0.8}H_{0.2}$ at ~29 K could be attributed to a superconductive transition. As the $\rho$-$T$ curves of annealed sample was measured on a pellet prepared by pressing the powders, the absence of zero resistivity is explained by effects of grain boundaries in the sample.

The electrical and magnetic measurements indicate that superconductivity with a $T_c^{onset}$ of 29 K is achieved by annealing the $CaFeAsF_{0.8}H_{0.2}$ at 553 K over 72 h under a He flow (99.999%, 200 mL/min). This $T_c^{onset}$ is higher than that achieved by direct electron doping of CaFeAsF *via* Co/Ni-substitution (25 K).[24-26] As no clear differences were observed in the compositions of the as-prepared and annealed samples, the origin of the superconductivity in the annealed sample can be explained by electron doping from the formation of fluorine vacancies, arising from selective elimination of hydrogen ($CaFeAsF_{1-x}H_x \rightarrow CaFeAsF_{1-x} + xe^- + 1/2xH_2^\uparrow$). Furthermore, the enhancement of $T_c$ compared with Co/Ni-substituted CaFeAsF may be explained by the indirect doping mode, and the lattice expansion along the *c* axis.

In summary, we have developed an indirect doping technique for *Ae*FeAsF and successfully induced bulk superconductivity with $T_c$ = 29 K in $CaFeAsF_{0.8}H_{0.2}$ by selective hydrogen elimination at 553 K for 72 h under a He flow (99.999%, 200 mL/min). The *c*-axis of the resulting superconducting sample is larger than that of CaFeAsF by 0.8%. The origin of the superconductivity can be explained by indirect electron doping from the formation of fluorine vacancies, which result from the selective elimination of hydrogen described by $CaFeAsF_{1-x}H_x \rightarrow CaFeAsF_{1-x} + xe^- + 1/2xH_2^\uparrow$.

**Acknowledgments**

This work was supported by the Funding Program for World-Leading Innovative R&D on Science and Technology (FIRST), Japan and the Element Strategy Initiative Project of the Ministry of Education, Culture, Sports, Science and Technology of Japan, T.H is a research fellow of Japan Society for the Promotion of Science (JSPS)

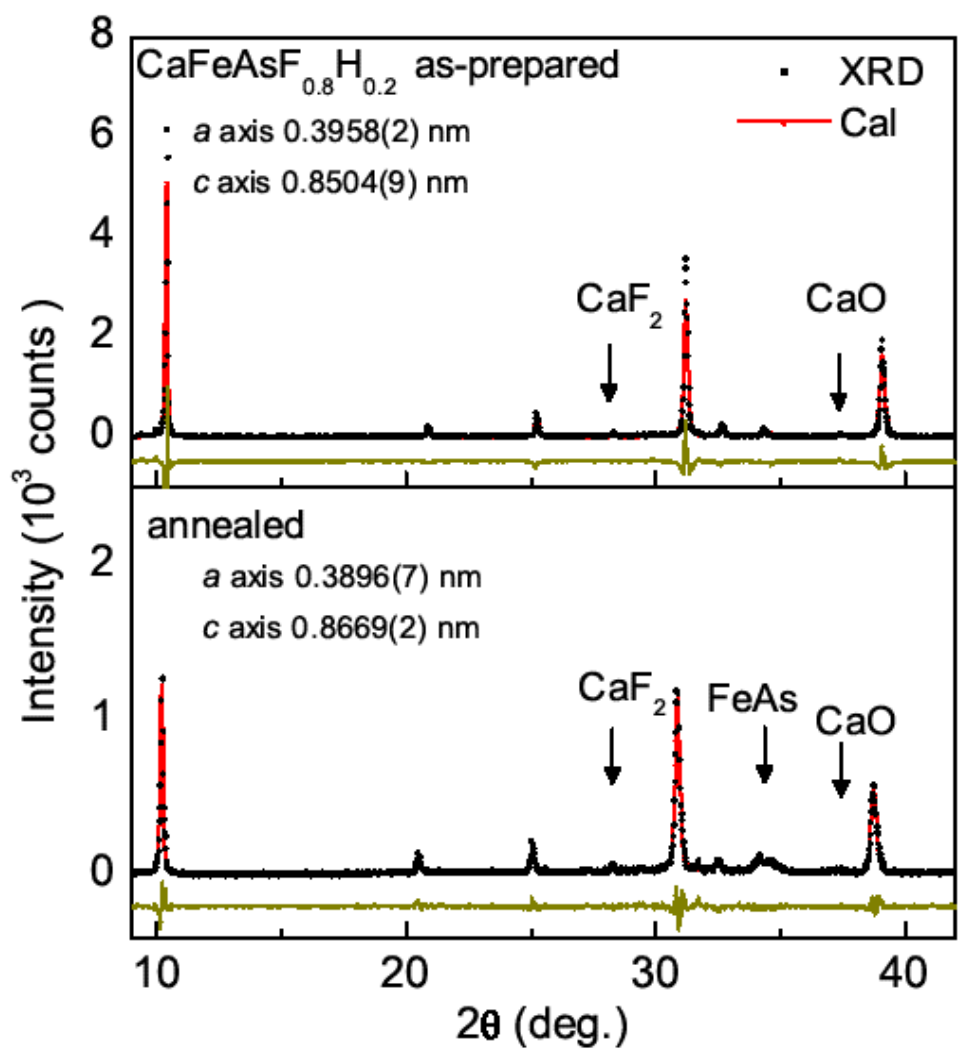

FIG.1. (a) Powder XRD patterns of as-prepared (top) and annealed (bottom) CaFeAsF$_{0.8}$H$_{0.2}$ at room temperature.

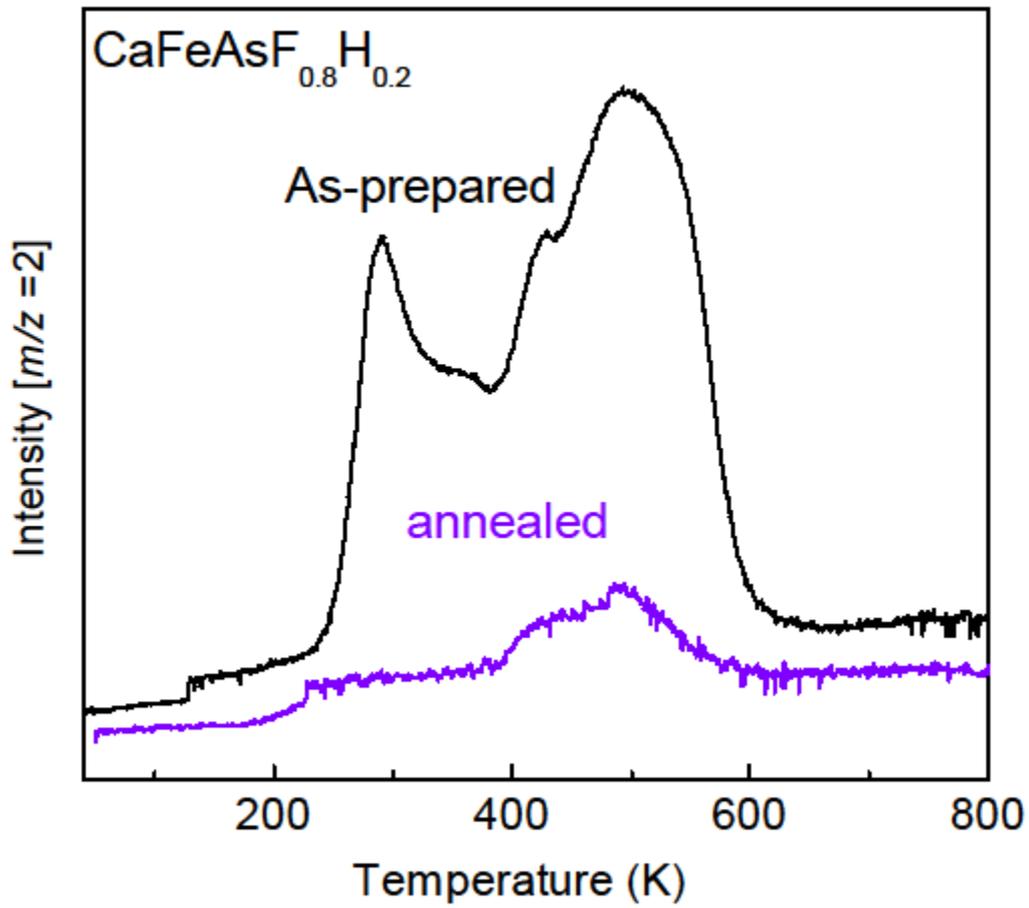

FIG.2. Integrated intensity of *m/z* = 2 MS trace *vs.* annealing temperature for as-prepared and annealed $CaFeAsF_{0.8}H_{0.2}$ samples measured using the TG-MS apparatus.

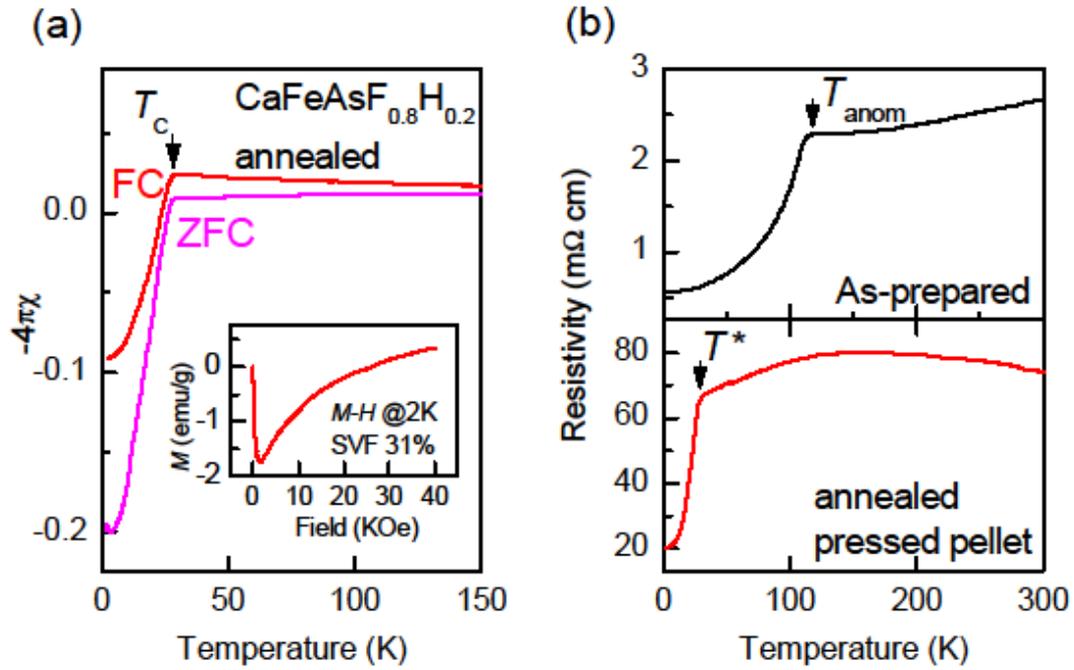

FIG.3. Electromagnetic properties of CaFeAsF$_{0.8}$H$_{0.2}$ (a) $\chi$-$T$ curves of annealed sample at 1Oe. Inset figure is $M$-$H$ curves of the sample. (b) $\rho$-$T$ curves of as-prepared and the annealed power compacts.

TABLE I. Element composition in CaFeAsF$_{1-x}$H$_x$ determined by EPMA and TDS.

| nominal $x$ | Ca$_{(EPMA)}$ | Fe$_{(EPMA)}$ | As$_{(EPMA)}$ | F$_{(EPMA)}$ | H$_{(TG-MS)}$ | O$_{(EPMA)}$ |
|---|---|---|---|---|---|---|
| CaFeAsF$_{0.8}$H$_{0.2}$ | | | | | | |
| As prepared | 1.00(1) | 1.00(1) | 0.95(1) | 0.73(1) | 0.24 | 0.07(3) |
| Annealed | 0.99(1) | 1.00(0) | 0.96(0) | 0.69(3) | 0.03 | 0.06(0) |